\documentclass[sigconf]{acmart}

\usepackage{booktabs}
\usepackage{caption}
\usepackage{balance}
\usepackage{tabularx}
\usepackage{multirow}
\usepackage[table]{xcolor} % beamer: \usepackage{colortbl}
\usepackage{array}
\usepackage{adjustbox}
\usepackage{makecell}
\definecolor{raiegray}{RGB}{217,217,217}
\newcommand{\raierow}{\rowcolor{raiegray}}
\usepackage{svg}
\newcolumntype{C}[1]{>{\centering\arraybackslash}m{#1}}
\renewcommand\arraystretch{1.15}
\usepackage{subcaption}
\usepackage{enumitem}
\usepackage{hyperref}
\usepackage{xurl}
\usepackage[ruled,vlined]{algorithm2e}

\SetAlFnt{\small}

\SetKw{Return}{return}

\AtBeginDocument{%
  }
\usepackage{graphicx}
\usepackage{subcaption}

\copyrightyear{2026}
\acmYear{2026}
\setcopyright{cc}
\setcctype{by}
\acmConference[WWW '26]{Proceedings of the ACM Web Conference 2026}{April 13--17, 2026}{Dubai, United Arab Emirates}
\acmBooktitle{Proceedings of the ACM Web Conference 2026 (WWW '26), April 13--17, 2026, Dubai, United Arab Emirates}
\acmPrice{}
\acmDOI{10.1145/3774904.3792451}
\acmISBN{979-8-4007-2307-0/2026/04}

\begin{document}

\title{RAIE: Region-Aware Incremental Preference Editing with LoRA for LLM-based Recommendation}

\author{Jin Zeng}
\authornote{Both authors contributed equally to this research.}
\email{zengj255@mail2.sysu.edu.cn}
\orcid{0009-0005-1468-6693}
\affiliation{%
  \institution{School of Cyber Science and Technology, Shenzhen Campus of Sun Yat-sen University}
  \city{Shenzhen}
  \state{Guangdong}
  \country{China}
}

\author{Yupeng Qi}
\authornotemark[1]
\email{qiyp7@mail2.sysu.edu.cn}
\orcid{0000-0002-6633-8616}
\affiliation{%
  \institution{School of Cyber Science and Technology, Shenzhen Campus of Sun Yat-sen University}
  \city{Shenzhen}
  \state{Guangdong}
  \country{China}
}

\author{Hui Li}
\email{hui@xmu.edu.cn}
\orcid{0000-0001-9139-3855}
\affiliation{%
  \institution{Department of Computer Science and Technology, Xiamen University}
  \city{Xiamen}
  \state{Fujian}
  \country{China}
}

\author{Chengming Li}
\email{licm@smbu.edu.cn}
\orcid{0000-0002-4592-3875}
\affiliation{%
  \institution{Artificial Intelligence Research Institute, Shenzhen MSU-BIT University}
  \city{Shenzhen}
  \state{Guangdong}
  \country{China}
}

\author{Ziyu Lyu}
\email{lvzy7@mail.sysu.edu.cn}
\authornote{Corresponding author.}
\orcid{0000-0002-3049-0085}
\affiliation{%
  \institution{School of Cyber Science and Technology, Shenzhen Campus of Sun Yat-sen University}
  \city{Shenzhen}
  \state{Guangdong}
  \country{China}
}

\author{Lixin Cui}
\email{cuilixin@cufe.edu.cn}
\orcid{0000-0003-1620-6532}
\affiliation{%
  \institution{School of Information, Central University of Finance and Economics}
  \state{Beijing}
  \country{China}
}

\author{Lu Bai}
\email{bailu@bnu.edu.cn}
\orcid{0000-0002-1033-8908}
\affiliation{%
  \institution{School of Artificial Intelligence, Beijing Normal University}
  \state{Beijing}
  \country{China}
}

\renewcommand{\shortauthors}{Jin Zeng et al.}

\begin{abstract}
Large language models (LLMs) are increasingly adopted as the backbone of recommender systems. However, user–item interactions in real-world scenarios are non-stationary, making preference drift over time inevitable. Existing model update strategies mainly rely on global fine-tuning or pointwise editing, but they face two fundamental challenges: (i) imbalanced update granularity, where global updates perturb behaviors unrelated to the target while pointwise edits fail to capture broader preference shifts; (ii) unstable incremental updates, where repeated edits interfere with prior adaptations, leading to catastrophic forgetting and inconsistent recommendations. To address these issues, we propose \textbf{R}egion-\textbf{A}ware \textbf{I}ncremental \textbf{E}diting (RAIE), a plug-in framework that freezes the backbone model and performs region-level updates. RAIE first constructs semantically coherent preference regions via spherical k-means in the representation space. It then assigns incoming sequences to regions via confidence-aware gating and performs three localized edit operations\textemdash Update, Expand, and Add\textemdash to dynamically revise the affected region. Each region is equipped with a dedicated Low-Rank Adaptation (LoRA) module, which is trained only on the region’s updated data.  During inference, RAIE routes each user sequence to its corresponding region and activates the region-specific adapter for prediction. Experiments on two benchmark datasets under a time-sliced protocol that segments data into Set-up (S), Finetune (F), and Test (T) show that RAIE significantly outperforms state-of-the-art baselines while effectively mitigating forgetting. These results demonstrate that region-aware editing offers an accurate and scalable mechanism for continual adaptation in dynamic recommendation scenarios. Our code is available at https://github.com/fengaogao/RAIE.
\end{abstract}




\maketitle
\newcommand{\userset}{\mathcal{U}}
\newcommand{\itemset}{\mathcal{V}}
\newcommand{\user}{u}
\newcommand{\itemm}{v}
\newcommand{\length}{L}
\newcommand{\sublength}{l}
\newcommand{\subwindowlength}{l_w}
\newcommand{\interhis}{S}
\newcommand {\sq}{A}
\newcommand{\llm}{\mathcal{M}}
\newcommand{\iindex}{i}
\newcommand{\konwindex}{k}
\newcommand{\targetindex}{\hat{k}}
\newcommand{\ts}{t_s}
\newcommand{\te}{t_e}
\newcommand{\tf}{t_f}
\newcommand{\ttime}{t}
\newcommand{\knowreg}{\Phi}
\newcommand{\prompt}{\mathbf{x}_u}
\newcommand{\hideden}{\mathbf{H}_u}
\newcommand{\ddim}{d}
\newcommand{\weight}{\mathbf{W}}
\newcommand{\lora}{\Delta}
\newcommand{\trainedlora}{\widehat{\delta}}
\newcommand{\ccenter}{c}
\newcommand{\ccenterset}{C}
\newcommand{\radius}{R}
\newcommand{\prototype}{{c}_k}
\newcommand{\K}{K}
\newcommand{\stride}{n}
\newcommand{\embedding}{\mathbf{e}}
\newcommand{\betsconfi}{p^\star}
\newcommand{\deltaa}{\delta}
\newcommand{\tauu}{\tau}
\newcommand{\gammaa}{\gamma}
\newcommand{\q}{q}

\newcommand{\ziyu}[1]{{\textcolor{red}{[ziyu: #1]}}}
\newcommand{\jin}[1]{{\textcolor{blue}{[zengjin: #1]}}}

\section{INTRODUCTION}
Recommender systems are widely deployed in online services such as e-commerce, video platforms and advertising to help users discover relevant content and reduce information overload~\cite{yang2022knowledge,zeng2025knowledge}. Among existing techniques, sequential recommendation models user behaviors as an ordered interaction sequence and predicts the next item by capturing temporal dependencies.  Representative methods include SASRec~\cite{kang2018self}, applying self-attention to model sequential patterns; TiSASRec~\cite{5e2d653a3a55acc837436820}, incorporating time-interval signals; and BERT4Rec \cite{sun2019bert4rec}, learning bidirectional dependencies via masked-item prediction. Despite their effectiveness, most sequential recommenders still rely on ID-based item representations, which encode limited semantic information and generalize poorly to unseen items or new domains.

Large language models (LLMs) alleviate the semantic limitations of ID-based recommenders~\cite{liu2023llmrec,geng2022recommendation} by leveraging item content to learn semantically grounded representations. Existing LLM-based recommenders can be broadly organized into three paradigms: (i) feature-based augmentation~\cite{rajput2023recommender}, where LLMs enrich user and item representations with semantic signals, (ii) generative recommendation~\cite{bao2023tallrec,luo2025recranker,lewis2020retrieval}, where LLMs are used to directly generate recommendation content via prompt, (iii) agentic recommenders~\cite{huang2025towards}, where LLMs act as agents that iteratively use tools and memory for recommendation. However, most prior work is trained and evaluated under static or near-stationary distributions. In practice, user interactions arrive continuously and data distributions evolve over time, leading to preference drift. When an LLM-based recommender is trained on historical data, it effectively learns a time-invariant mapping from context to relevance; as user interests shift, this mapping becomes stale, resulting in degraded recommendation performance. Therefore, dynamic recommendation requires drift-aware mechanisms that can stably adapt to evolving preferences.

\begin{figure}[t]
  \centering
  \includegraphics[width=\linewidth]{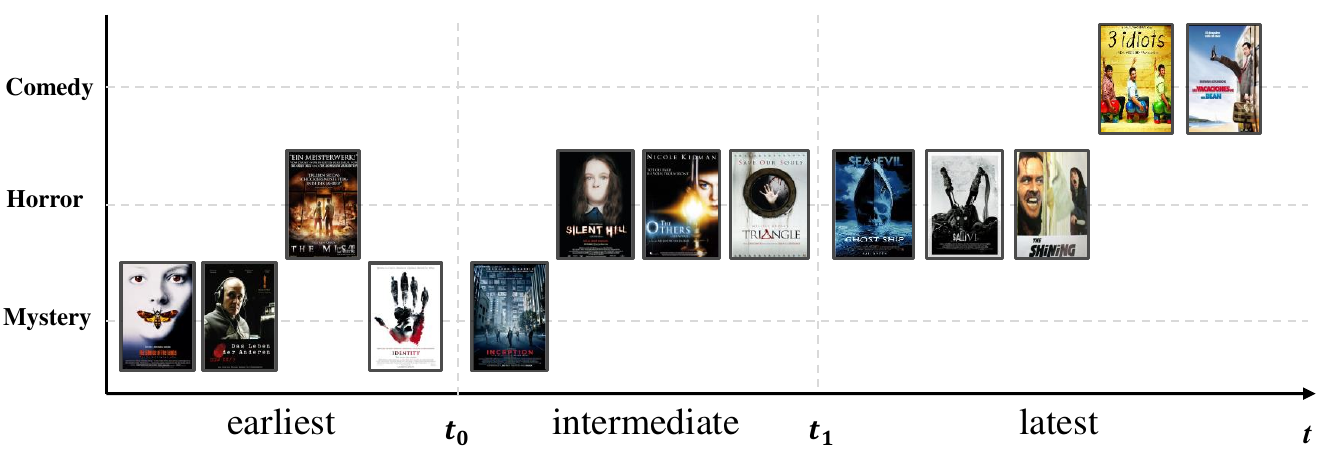}
  \caption {Illustrated example of user preference drift in the movie domain.}
  \label{example}
\end{figure}

Figure \ref{example} illustrates a preference-drift example. For clarity, we partition a user’s interactions into three consecutive time slices: earliest, intermediate and latest. In the earliest slice, the user mainly prefers \emph{mystery}; in the intermediate slice, the preference shifts toward \emph{horror}, which remains dominant in the latest slice, while the latest slice also shows a short-term increase in \emph{comedy}. If a static LLM trained only on the earliest slice is used to make recommendations for the latest slice, it will often overestimate the user’s interest in \emph{mystery}, failing to capture both the dominant preference and the transient shifts observed in the latest slice. A straightforward solution is periodic retraining, but it is computationally expensive and can overwrite valid long-term preferences under distribution shifts, leading to catastrophic forgetting~\cite{kirkpatrick2017overcoming,parisi2019continual}. Parameter-efficient fine-tuning (PEFT) offers a more efficient alternative by updating a small number of parameters while freezing the backbone. For example, LoRA updates low-rank matrices with low training and storage costs~\cite{hu2022lora}, and adapter-based methods also support efficient transfer and fast adaptation~\cite{houlsby2019parameter}. In practice, many PEFT pipelines rely on a single global adapter to handle distributional shifts. Although effective, such global updates may perturb stable preferences and introduce interference across behavioral patterns, resulting in a process that lacks specificity and precise control. Recent multi-adapter variants (e.g. dual-LoRA~\cite{shi2024preliminary} or mix of experts LoRA~\cite{wu2024mixture}) enhance adaptation capacity by assigning different adapters to different tasks. However, such methods often route data to adapters using task-level heuristics, and their parameter updates are weakly aligned with user preference changes, making it difficult to precisely adapt to the specific drifting regions.

Inspired by knowledge editing for LLMs~\cite{meng2022locating,meng2022mass,mitchell2022memory}, we propose RAIE, a plug-in, region-aware incremental editing framework for LLM-based recommendation to address preference drift. A key observation is that preference drift is often localized: a user may remain stable on some interests while drifting within others~\cite{wei2025stability}. This calls for a structured update scheme that groups interactions into interest groups and localizes adaptation to the interest group experiencing drift, thereby preserving stable preferences elsewhere. We therefore introduce knowledge regions, defined as semantically coherent clusters of a user’s historical interactions in a shared representation space, each summarized by a centroid and an effective radius. By organizing interactions into regions, RAIE can localize both routing and updates: when drift occurs, it updates only the affected region, reducing interference with stable regions. RAIE consists of three components: (a) \textbf{knowledge region construction}, which partitions the user’s historical interactions into multiple regions and assigns a LoRA adapter to each region; (b) \textbf{region-aware editing and LoRA adaptation}, which uses confidence-aware routing to trigger edit operations and trains the corresponding regional adapter on updated region data; and (c) \textbf{region-aware routing}, which activates the region-specific adapter at inference to reduce interference across regions. Focusing on component (b), RAIE supports three regional edit operations to track preference evolution: Update refines the region’s representation under fixed boundaries; Expand relaxes the boundary moderately and applies the same refinement to cover mild outward shifts; Add instantiates a new region when a novel pattern emerges. Refinement uses exponential-moving-average (EMA) updates for stability. A similarity-based router assigns new sequences to candidate regions, and a confidence threshold determines whether an edit is triggered. Each knowledge region is paired with a LoRA adapter initialized uniformly. After editing, the regional data are used to train its adapter, aligning the adapter’s parameters with the region’s current sub-distribution.

In summary, our contributions are highlighted as follows:
\begin{itemize}[leftmargin=*]
\item We formalize user preference drift adaptation in LLM-based recommendation as a region-aware incremental editing problem, highlighting the core trade-off between localized adaptation and global stability. This formulation provides a structured approach to manage evolving user preferences while maintaining model coherence.
\item We introduce the Region-Aware Incremental
Editing (RAIE) framework, which integrates three key components: Knowledge Region Construction, Region-Aware Preference Editing and LoRA
Adaptation, while reducing cross-region interference via region-specific adapters and routing.
\item Extensive experiments on MovieLens-10M and Yelp demonstrate that RAIE consistently outperforms state-of-the-art baselines, while effectively adapting to emerging user preferences and mitigating forgetting in cross-phase incremental learning scenarios.
\end{itemize}

\section{RELATED WORK}
\subsection{Continual learning}
User preferences are inherently non-stationary, evolving over time due to changing contexts. This preference drift causes models trained on static historical data to become misaligned with current user interests~\cite{koren2009collaborative}. While periodic full retraining of conventional sequential recommenders (e.g., SASRec~\cite{kang2018self}, BERT4Rec~\cite{sun2019bert4rec}) is common, it becomes prohibitively costly for LLM-based backbones due to massive computational and storage overhead. Moreover, retraining from scratch can exacerbate catastrophic forgetting, disrupting previously learned stable preferences. Continual learning offers a framework to balance stability (retaining old knowledge) and plasticity (acquiring new knowledge)~\cite{yoo2025continual}. Representative strategies include: (i) regularization-based methods (e.g., EWC~\cite{kirkpatrick2017overcoming}, $\ell_2$-SP\cite{xuhong2018explicit}) that constrain parameter updates to protect important weights, and (ii) replay-based methods~\cite{shin2017continual,klasson2022learn} that revisit past data to mitigate forgetting. Despite these advances, applying continual learning to LLM-based recommenders remains challenging in practice. Regularization-based methods still require repeated gradient updates on a large backbone, while replay-based methods incur non-trivial storage and computation overhead and can amplify training instability as updates accumulate.

\subsection{Parameter-Efficient Fine-Tuning for LLM-based Recommendation systems}
To enable feasible adaptation of large models, Parameter-Efficient Fine-Tuning (PEFT) has emerged as a key paradigm~\cite{ding2023parameter}. Techniques like low-rank adaptation (LoRA)~\cite{hu2022lora}, adapters~\cite{houlsby2019parameter} and prefix tuning~\cite{li2021prefix} freeze the pre-trained backbone and only optimize a small set of injected parameters, drastically reducing update costs. To enhance capacity and specialization, recent work introduces routing mechanisms. For instance, instance-level LoRA (iLoRA~\cite{kong2024customizing}) and Mixture of LoRA Experts (MoLE~\cite{wu2024mixture}) dynamically select or combine multiple adapters to tailor the model’s behavior. These PEFT and routing techniques are increasingly integrated into modern LLM-for-Rec pipelines~\cite{xu2024openp5,rajput2023recommender}. However, adapter selection is often driven by task-level heuristics, and the resulting updates are not explicitly tied to how user interests evolve over time, which makes it hard to apply targeted adaptations without introducing unnecessary interference.

\subsection{Knowledge Editing for LLMs and Recommenders}
Knowledge editing aims to make precise, localized changes to a model's behavior without full retraining~\cite{zhang2024comprehensive}. Parametric editing methods (e.g., ROME~\cite{meng2022locating} and MEMIT~\cite{meng2022mass}) directly modify specific model weights to alter factual associations, while external memory-based methods (SERAC~\cite{mitchell2022memory}, MEND~\cite{mitchell2021fast}) store edits in a side network to override the base model's predictions. In recommendation, editing concepts have been applied to correct user-item interactions~\cite{lai2024better} or update knowledge graphs~\cite{rozner2024knowledge}. Lifelong editing frameworks like ELDER~\cite{li2025elder} explore LoRA-based routing to manage sequential edits. Nevertheless, most work remains focused on editing factual knowledge or discrete interaction records, rather than modeling and editing continuous preference subspaces. We advocate a preference-oriented editing perspective, whose central goal is to calibrate the user's dynamically evolving interests. Our proposed RAIE framework instantiates this perspective: it organizes user behaviors into semantically coherent regions, treats each preference region as an editable unit, assigns it a dedicated LoRA module, and strictly localizes updates to regions experiencing drift, thereby bridging the gap between knowledge editing and preference adaptation in recommendation.
\section{Problem Definition}
\textbf{Interaction Sequences and Temporal Split.} In the  recommendation scenario, we have a user set $\userset$  and an item set $\itemset$. 
For a user $\user_i \in\userset$, their time-ordered interaction history is denoted as $\interhis_{\user_i}=[\itemm_{i,1},\itemm_{i,2},\ldots,\itemm_{i,\length_{\user_i}}]$, where each item $\itemm_{i,j}$ is associated with a timestamp $t_{i,j}$.
In order to perform incremental learning and updates, we split each user's sequence into three temporal phases:
$\interhis^S$ (Set-up phase), $\interhis^F$ (Incremental Fine-tune phase), and $\interhis^T$ (Inference Test phase). Specifically, $\interhis^S$ contains interactions with $t \le t^S$, $\interhis^F$ contains those with $t^S < t \le t^F$, and $\interhis^T$ contains the remaining interactions with $t > t^F$.

\textbf{Task Definition.}
The incremental recommendation process consists of three stages: set‑up phase, dynamic fine‑tuning phase on incoming sequences, and final inference test phase. As time progresses, new data splits are continuously generated. The system works dynamically by performing incremental fine-tuning on each new training splits, followed by inference testing on the corresponding test splits. For simplicity, we state the problem formulation with the three phases defined above, i.e., $\interhis^S$, $\interhis^F$, $\interhis^T$.
\paragraph{Set-up Phase (S)} We construct the initial model using the early-stage user sequences $\interhis^S$.  We adopt a base recommendation model $\llm(\theta)$, which can be traditional sequential recommendation methods or LLM-based methods\footnote{In the method section, we mainly introduce that the sequence representation extraction from LLM-based recommendation methods. But our proposed framework is plug-in method that is flexible for traditional recommendation methods. We present experimental results with diverse base models to demonstrate this generality.}. This model is used to obtain initial user sequence representations from $\interhis^S$. For LLM-based recommenders, we define a prompt construction function $\mathrm{BuildPrompt}(\cdot)$ that maps a user subsequence to a textual prompt\footnote{Prompt template: Here is the purchase history of user\_\{user\_id\}: item \{history\}. I wonder what is the next recommended item for the user. Answer:}. In addition, we construct a set of initial preference regions $\knowreg={(\ccenter_k,\radius_k)}_{k=1}^{K}$, where $K$ is the initial number of regions, $\ccenter_k$ is the center of region $k$, and $\radius_k$ defines its effective radius in the representation space.
 
\paragraph{Incremental Learning Finetune Phase (F)} This is the core phase of incremental recommendation. Its primary goal is to discover the fine-grained preference changes within the existing preference regions $\knowreg$ and to perform incremental preference updating to the recommendation model $\llm$ (i.e., $\llm(\theta) \rightarrow \llm(\theta')$) based on the new user sequences $\interhis^F$.
\paragraph{Inference Test Phase (T)} The phase is to perform user preference prediction. Based on the updated model $\llm(\theta^{'})$, the new preference regions $\knowreg$, and the test user sequence $\interhis^T$, the model performs next-item prediction for each test sequence to evaluate the effectiveness of the incremental adaptation.

\begin{figure*}[t]
\centering
\includegraphics[width= \textwidth]{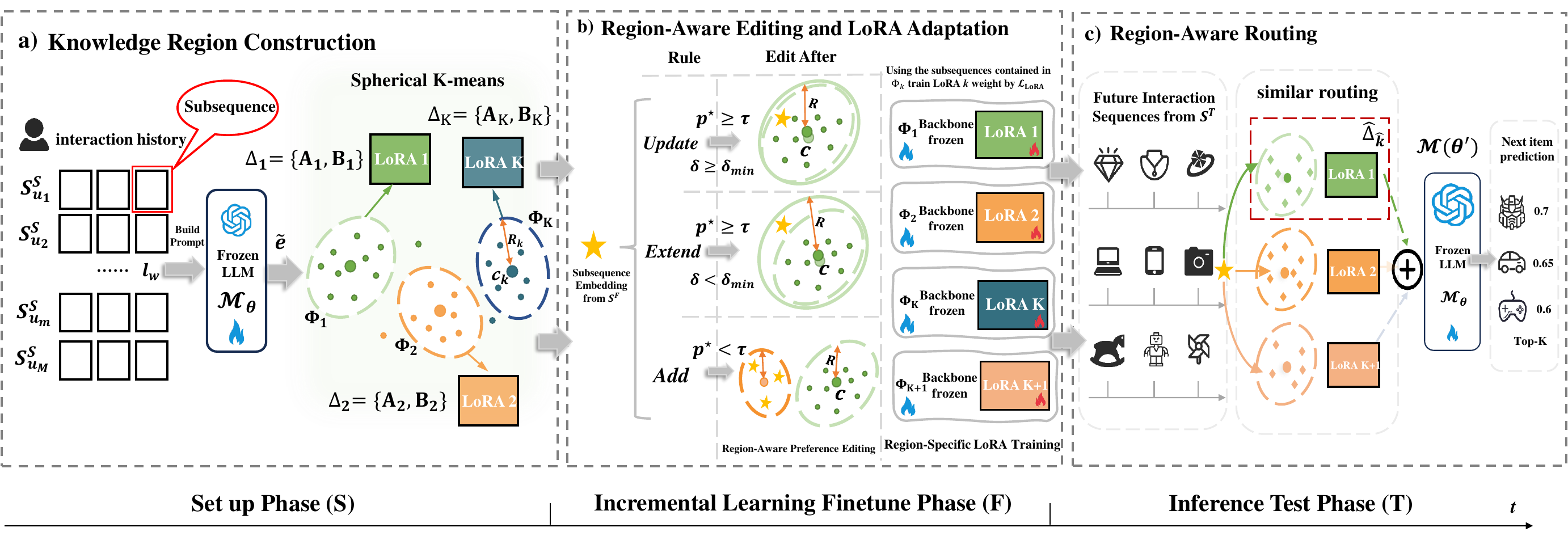}
\caption {The overall architecture of our proposed RAIE. It consists of three core modules, respectively. a) Knowledge Region Construction at the set-up phase (S); b) Region-aware Editing and LoRA Adaption in the incremental learning finetune phase (F); and the Region-ware Routing for the inference test phase (T).}
\label{fig1}
\end{figure*}

\section{Method}
To address the challenges of incremental preference updating, we propose Region-Aware Incremental Editing (RAIE), a framework that integrates low-rank adaptation for LLM-based recommendation. The overall architecture of RAIE is shown in Figure~\ref{fig1}. Our method consists of three key components: (1) Knowledge Region Construction, (2) Region-Aware Editing and LoRA Adaptation, and (3) Region-Aware Routing. RAIE operates dynamically across three phases: set-up (S), incremental fine-tuning (F), and inference (T). During the set-up phase, initial knowledge regions are constructed, each paired with a LoRA-based parameter-efficient module. The region-aware editing and LoRA adaptation module serves as the core mechanism in the incremental fine-tuning phase. It locates the corresponding region for each new preference sequence and performs incremental editing through three operations: Update, Expand, and Add. Based on the edited regions, we finetune the corresponding LoRA adapters to enable efficient incremental learning. During inference, the region-aware routing mechanism activates the appropriate adapter for next-item prediction. Algorithm~\ref{alg:raie} summarizes the overall procedure of method.

\subsection{Knowledge Region Construction}
\label{sec:RegionConstuct}
We perform knowledge region construction using user sequence $\interhis^s$ from the set-up phase. This process consists of three steps: (1) segmenting user sequences into subsequences and extracting their representations, (2) discovering preference regions via clustering, and (3) mapping each region to a dedicated LoRA adapter.
\subsubsection{Subsequence Segmentation and representation}
\label{sec:segment}

Since user sequences are often long and contain multiple intents over time, capturing fine-grained preferences directly from the full sequence is difficult. Therefore, we first segment each user sequence into a series of overlapped subsequences and obtain their representations. Specifically, we adopt a sliding window approach~\cite{Kang2018SASRec} with window length $\subwindowlength$ and stride $\stride$. For the user sequences $\interhis^s$, this yields subsequences $[\sq_1,\sq_2,\ldots,\sq_{\lfloor (T_u-L_w)/\stride \rfloor+1}]$.\footnote{The same segmentation is applied to sequences in the finetuning and inference phases for consistent processing.} This fragment-level view emphasizes temporal locality and isolates short-term intents, highlighting fine-grained user preferences.

After subsequence segmentation, we extract the embedding representation for each subsequence using an LLM-basd backbone. We construct the user preference prompt by filing the user subsequence into a template: $\prompt=\mathrm{BuildPrompt}(\sq)$ and obtain the embedding representation from the frozen backbone $\llm(\prompt)$. We take the last-token hidden state of the last hidden layer as the subsequence preference representation $\embedding\in\mathbb{R}^d$. For further region clustering in the spherical space, we apply $\ell_2$-normalization on the preference represenation $\embedding$, and finally obtain the normalized representation $\tilde\embedding=\embedding/\|\embedding\|_2$.

\subsubsection{Preference Region Discovery}
We treat all the subsequences from all users in the set-up phase as clustering samples, and apply the spherical $k$-means method~\cite{dhillon:modha:mlj01} on the preference representation of each sample to perform preference region discovery. The spherical $k$-means method will cluster the similar preference into a preference region, with the intuition that the samples within a cluster share similar preferences, and the samples from difference clusters have different preferences. The optimization objective is defined as in Equation~\ref{eqn:skmeans}, the initial number of regions is set as k.  
\begin{equation}
\label{eqn:skmeans}
\max_{\{\ccenterset_\konwindex,\prototype\}_{\konwindex=1}^\K}\ \sum_{\konwindex=1}^{\K}\ \sum_{\tilde\embedding\in \ccenterset_\konwindex}\ \prototype^\top  \tilde{\embedding}
\quad \text{s.t.}\quad \|\prototype\|_2=1.
\end{equation}
where $\ccenterset_\konwindex$ denote the set of samples for each preference region, $\prototype$ is center prototype  of each cluster region.

When the clustering algorithm finishes, we obtain K preference regions $\knowreg=\{(\ccenter_\konwindex,\radius_\konwindex)\}_{\konwindex=1}^{\K}$. Each region is characterized by a prototype representation $\prototype$ and a radius $\radius_k$. The radius quantifies the intra-region cohesion, parameterizing coverage and locality.

\begin{algorithm}[ht]
\caption{Region-Aware Incremental Editing}
\label{alg:raie}
\SetAlgoLined
\LinesNumbered
\DontPrintSemicolon
\SetKwInOut{Input}{\textbf{Input}} 
\SetKwInOut{Output}{\textbf{Output}} 

\Input{User set $\userset$; frozen encoder $\llm_\theta$; data splits $\interhis^{S}$, $\interhis^{F}$, $\interhis^{T}$}
\Output{ top-k item recommendation list}

\For{each $\interhis^{S}_{u} \in \interhis^S$}{
    Segment user sequence $\interhis^S_u$ to obtain subsequences $\sq$\;
    
    Build textual prompt $\prompt$ and encode it using $\llm{(\theta)}$ to obtain $\tilde\embedding$\;
}

\tcp{1. Knowledge Region Discovery}

Apply spherical $k$-means clustering to obtain centers $\ccenter$ and radius $\radius$\;

Form knowledge region set $\knowreg=\{(\ccenter_k,\radius_k)\}_{\konwindex=1}^{\K}$\;

\tcp{2. Region-Aware Preference Editing}

\For{each incoming subsequence $\interhis^{F}_{u} \in \interhis^F$}{
    perform subsequence segment and obtain the vector $\tilde\embedding$\;
    
    Select candidate region $\hat{\knowreg_k}$ via cosine similarity Eq.~(\ref{eq:simdef})\;
    
    \uIf{$\betsconfi\ge \tauu \ \text{and}\ \deltaa\ge \deltaa_{\min}$}{
        Update region center and radius via Eqs.~(5)\;
    }
    \uElseIf{$\betsconfi\ge \tauu \ \text{and}\ \deltaa< \deltaa_{\min}$}{
        Expand boundary using Eqs.~(\ref{eq:expansion_radius})
    }
    \Else{
        Create new region by Eqs.(1)
    } 
}

\tcp{3. Region-Specific LoRA Training}

\For{each region $\konwindex=1,\ldots,\K$}{
    Insert low-rank adapters $\Delta \theta_{\konwindex}$\;
    
    Train $\Delta \theta_\konwindex$ on $\interhis_\konwindex=\interhis_\konwindex^{(S)}\cup\interhis_\konwindex^{(F)}$ using next-item loss Eq.~(\ref{eq:next})
}
Activate trained LoRA
adapters and score candidates via Eq.(10)

\Return{top-k item recommendation list}
\end{algorithm}

\subsubsection{Regional LoRA Mapping}
After obtaining the preference regions, we aim to conduct region-aware incremental learning for the recommender model. We adopt a region-specific LoRA strategy: each preference region is mapped to a dedicated LoRA adapter, which is updated in a parameter-efficient manner to capture localized preference knowledge during the $F$ phase. Here we demonstrate the regional LoRA mapping mechanism. 

In general, LoRA-based parameter-efficient learning for a recommender model is defined as:
\begin{equation}
    \max_{\Delta \theta }  \sum_{(\prompt, \itemm) \in \interhis} \sum_{t=1}^{L}
    {\log \llm_{\theta+\Delta \theta}(\itemm_t|\prompt, \itemm_{<t} )}
    \label{eqn:LoRA}
\end{equation}
where $\Delta \theta$ is decomposed into two low-rank matrices $\mathbf{A}\in\mathbb{R}^{r\times d_{\text{in}}}$,
$\mathbf{B}\in\mathbb{R}^{d_{\text{out}}\times r}$.  When performing finetuning or incremental learning on the new data, we only update the low-rank matrices, instead of the full set of model parameters, ensuring parameter efficiency.

In order to perform the fine-grained and localized preference updating, we construct the region-specific LoRA, and initially map each preference regions with a region specific LoRA. Namely, for each preference region $\knowreg_k$ is associated with region specific LoRA matrices $\mathbf{A}^k$ and $\mathbf{B}^k$.  In the incremental learning phase, as the number of preference regions increases, a new region-specific LoRA is constructed. Incremental learning then proceeds in a parameter-efficient manner, whereby each region is updated via its corresponding LoRA module as defined in Equation ~\ref{eqn:LoRA}.

\subsection{Region-Aware Preference Editing and LoRA Adaptation}
After the set-up phase, the core part of our method is to perform region-aware preference editing and incremental learning via LoRA adaption, using the new user sequence $\interhis^F$.

\subsubsection{Region-Aware Preference Editing}

\paragraph{Region-aware Preference Region Localization} 
Before preference editing, the preliminary process is to determine which regions to edit. For a new user preference subsequence $\sq^F$ in the $F$ phase, we have the preference representation $\tilde{\embedding}$ (same operation as in Section~\ref{sec:RegionConstuct}). We then calculate the similarities between the preference represenation and each region’s center representation:

\begin{equation}
\label{eq:simdef}
Score_\konwindex=\ccenter^\top_\konwindex \tilde\embedding,\quad
p_\konwindex=\frac{\exp(Score_\konwindex)}{\sum_{j=1}^\K \exp(Score_j)}
\end{equation}
where $Score_k$ denotes the similarity score between the new preference representation with a region center, $p_\konwindex$ indicates the confidence score. We can  sort the regions based on the confidence scores, and choose the preference regions with the highest confidence scores as the candidate matching regions.

\paragraph{Preference Region Editing Mechanism}
When getting the candidate matching regions, we define some rules to select the editing operation for each preference region as in Equation~\ref{eqn:updatingrule}. We denote $\betsconfi=p_{\hat{\konwindex}}$ be the highest confidence, and $\delta=\betsconfi-\max_{j\neq \hat{\konwindex}}p_j$ be the upper confidence margin between the highest confidence score and the second highest confidence score. If the top confidence is smaller than a threshold $\tauu$, it indicates there is no similar region and requires to add a new region with the \textbf{Add} operation. But if the top confidence is larger than a threshold $\tauu$, it indicates there are some similar regions, perform the \textbf{Update} operation or the \textbf{Expand} operation on the matching regions. If upper confidence margin is large ($\delta_{\min}>0$ is the minimum margin), we should \textbf{Update} the matching region, otherwise perform the \textbf{Expand} operation.
\begingroup
\small
\renewcommand{\arraystretch}{0.95}
\begin{equation}
\label{eqn:updatingrule}
\mathrm{Act}(\tilde\embedding)=
\begin{cases}
%\texttt{add},     & \text{if } \mathbf{c}_{\hat{k}}^\top \embedding<\sigma \ \text{and}\ \betsconfi<\eta\,\tauu,\\
\texttt{update},  & \text{if } \betsconfi\ge \tauu \ \text{and}\ \deltaa\ge \deltaa_{\min},\\
\texttt{expand},  & \text{if } \betsconfi\ge \tauu \ \text{and}\ \deltaa< \deltaa_{\min},\\
\texttt{add},     & \text{if} \betsconfi < \tauu
\end{cases}
\end{equation}
\endgroup

The three editing operations are defined as follows:
\begin{itemize}[leftmargin=*]
\item \textbf{Update:} updating the center and radius as follows:
\begin{equation}
\label{eq:update_radius}
\radius_{\hat{\konwindex}}\leftarrow (1-\beta)\radius_{\hat{\konwindex}}+\beta\arccos(\mathbf{c}_{\hat{\konwindex}}^\top \tilde\embedding) \\ \nonumber
\end{equation}
\begin{equation}
\label{eq:update_center}
\tilde{\mathbf{c}}_{\hat{\konwindex}}=(1-\gamma)\mathbf{c}_{\hat{\konwindex}}+\gamma \tilde\embedding,\quad
\mathbf{c}_{\hat{\konwindex}}\leftarrow\frac{\tilde{\mathbf{c}}_{\hat{\konwindex}}}{\|\tilde{\mathbf{c}}_{\hat{\konwindex}}\|_2}.
\end{equation}
where $\beta$ and $\gamma$ are smoothing coefficients.

\item \textbf{Expand:} Expand the boundary of the preference regions while keeping semantically aligned:
\begin{equation}
\label{eq:expansion_radius}
\radius_{\hat{\konwindex}}\leftarrow \radius_{\hat{\konwindex}}+\lambda\big(\arccos(\mathbf{c}_{\hat{\konwindex}}^\top \tilde\embedding)-\radius_{\hat{\konwindex}}\big)_+,
\quad R_{\hat{\konwindex}}\le \radius_{\max} \nonumber
\end{equation}
\begin{equation}
\label{eq:expansion_center}
\mathbf{c}_{\hat{\konwindex}}\leftarrow
\frac{(1-\alpha)\mathbf{c}_{\hat{\konwindex}}+\alpha\,\,\tilde\embedding}
     {\|(1-\alpha)\mathbf{c}_{\hat{\konwindex}}+\alpha\,\,\tilde\embedding\|_2}.
\end{equation}
where $\lambda>0$ controls the expansion rate, $\alpha \in (0,1)$ and $\radius_{\max}$ caps the maximum radius.

\item \textbf{Add} : When the exsiting regions are not similar with the new user preference, we need to create a new region via the add operation.

We design the bath-level new region construction. Namely, we construct a buffer pool $\mathcal{B}$  and fill the new user preference sample into the buffer pool when it triggers the \textbf{Add} operation. When $|\mathcal{B}_{\mathrm{add}}|$ exceeds a threshold, we perform spherical $\konwindex$-means on $\mathcal{B}$ as defined in Equation~\ref{eqn:skmeans}.
\end{itemize}

\subsubsection{Region-Specific LoRA Training}
After performing preference editing, we train the corresponding region-specific LoRA adapters. As mentioned above, each preference region $\knowreg_k$ corresponding to a Region-Specific LoRA with two low-rank matrices $\mathbf{A}^k$ and $\mathbf{B}^k$. We perform Region-Specific LoRA Training on $\interhis_\konwindex=\interhis_\konwindex^{(S)} \cup \interhis_\konwindex^{(F)}$ by optimizing the following loss function $\mathcal{L}=\mathcal{L}_{\text{LoRA}}+\mathcal{L}_{p}$:

\begin{equation}
\label{eq:next}
\mathcal{L}_{\text{LoRa}}(\Delta \theta_k ;\interhis_\konwindex)
=  -\sum_{(\prompt, \itemm) \in \interhis_k} \sum_{t=1}^{L}
    {\log \llm_{\theta+\Delta\theta_{k}}(\itemm_t|\prompt, \itemm_{<t} )}
\end{equation}

\begin{equation}
\label{eq:l_sep}
\mathrm{d}_{i,j}=\big(\radius_i+\radius_j-\|\mathbf{c}_i-\mathbf{c}_j\|_2\big)_+,\qquad
\mathcal{L}_{p}=\lambda_{\mathrm{sep}}\sum_{i<j}{\mathrm{d}_{i,j}}^2.
\end{equation}
where $\mathcal{L}_{\text{LoRA}}$ is the LoRA training loss and $\mathcal{L}_p$ is the a penalty term, which control the potential ovelap between different regions.

\subsection{Region-Aware Routing.}
In the inference phase, we have the trained region-specific LoRA adapters
$\{\widehat{\lora}_k\}_{k=1}^{K}$. Given a new user subsequence, we obtain its preference representation
$\tilde\embedding$ and select the most compatible preference region $\knowreg_{\hat{k}}$ based on the similarity:
\begin{equation}
\knowreg_{\hat{k}}=\operatorname*{arg\,max}_{\konwindex\in\{1,\ldots,\konwindex\}}\ \tilde\embedding^{\top}\ccenter_\konwindex.
\end{equation}
We then attach the corresponding adapter $\widehat{\lora}_{\hat{\konwindex}}$ to the frozen backbone
$\mathcal{M}_{\Theta}$. Let $f(\itemm\mid \prompt, \mathcal{M}_{\Theta},\widehat{\lora}_{\hat{\konwindex}})$ denote the
routed model’s logit for item $\itemm$. The predictive distribution over the candidate set
$\itemset$ is
\begin{equation}
p(\itemm\mid \prompt,\llm,\widehat{\lora}_{\hat{\konwindex}})=
\frac{\exp\!\big(f(\itemm\mid \prompt,\llm,\widehat{\lora}_{\hat{\konwindex}})\big)}
     {\sum_{j\in \itemset}\exp\!\big(f(j\mid\prompt,\llm,\widehat{\lora}_{\hat{\konwindex}})\big)}.
\end{equation}
The top-$k$ items under this distribution are returned as the final recommendation list.

\section{EXPERIMENTS}
To evaluate the performance of RAIE (a plugin-in module for existing recommenders), we conduct experiments to address the following research questions:

\noindent
\textbf{• RQ1}: How does RAIE compare with state-of-the-art plug-in baselines across different backbone models and datasets?

\noindent
\textbf{• RQ2}: What is the contribution of each key component to overall performance?

\noindent
\textbf{• RQ3}: How sensitive is RAIE to major hyperparameters, and are the trends consistent across datasets?

\noindent
\textbf{• RQ4}: Does RAIE provide intuitive interpretability reflecting pre-edit and post-edit changes in preference structure?

\begin{table}[h]
\centering
\caption{Statistics of the datasets under the S$\rightarrow$F$\rightarrow$T protocol.}
\label{tab:datasets_grouped}
\setlength{\tabcolsep}{5pt}      % narrower columns
\scriptsize                      % smaller font
\begin{tabular}{llrr}
\toprule
\textbf{Stage} & \textbf{Metric} & \textbf{MovieLens-10M} & \textbf{Yelp} \\
\midrule
\multirow{3}{*}{Set-up (S)} 
  & \#Users         & 43,021    & 41,072 \\
  & \#Items         & 4,784     & 28,404 \\
  & \#Interactions  & 2,502,840 & 702,212 \\
\midrule
\multirow{3}{*}{Finetune (F)} 
  & \#Users         & 18,168    & 39,631 \\
  & \#Items         & 8,305     & 29,657 \\
  & \#Interactions  & 1,501,707 & 421,328 \\
\midrule
\multirow{3}{*}{Test (T)} 
  & \#Users         & 14,450    & 30,274 \\
  & \#Items         & 9,264     & 27,437 \\
  & \#Interactions  & 1,001,137 & 280,885 \\
\bottomrule
\end{tabular}
\end{table}

\subsubsection{Datasets.}
We evaluate on two public datasets: \textbf{MovieLens-10M}\footnote{\url{https://grouplens.org/datasets/movielens/10m/}}and \textbf{Yelp}\footnote{\url{https://www.kaggle.com/datasets/yelp-dataset/yelp-dataset}} (statistics in Table~\ref{tab:datasets_grouped}). Following common practice for implicit recommendation, we binarize feedback by marking ratings $\ge 4$ as positive and discarding the rest~\cite{lin2023self}.
To reduce sparsity and noise, we apply $k$-core filtering (remove users and items with fewer than $k$ interactions): $k{=}5$ for MovieLens-10M and $k{=}10$ for Yelp. Then, we sorted the interaction records of each user in chronological order to obtain the item sequences. To enforce a consistent temporal protocol, we first compute two dataset-level timestamp quantiles, $q^{S}=0.5$ and $q^{F}=0.8$, and map them to cutoffs $t^S$ and $t^F$. For each user, the chronologically ordered interactions are then split into three disjoint segments: Set-up (S) with $\mathrm{t}<t^S$, Finetune (F) with $t^S \le \mathrm{t} < t^F$, and Test (T) with $\mathrm{t}\ge t^F$. Within each segment, we form training examples via right-aligned sliding windows: the most recent items serve as context and the immediate next item from the same segment is the prediction target; any window that would cross a segment boundary is discarded to prevent leakage. 
\subsubsection{Implementation details.}
We use AdamW~\cite{loshchilov2017decoupled} with learning rate $5\times10^{-4}$. Models are trained for 5 epochs in the set-up phase and 3 epochs in the fine-tuning phase. LoRA employs rank $r=8$, scaling factor $\alpha=16$, and dropout $0.05$. For region construction, spherical $k$-means uses radius quantile $q$=0.9; region-specific adapters are trained for 3 epochs with an original-data mixing ratio of $0.7$.

\begin{table*}[t]
  \centering
\caption{Results on MovieLens-10M (left) and Yelp (right) under the Set-up (S) and Test (T) splits, reporting Recall@10 (R@10) and NDCG@10 (N@10). The Set-up (S) split evaluates knowledge retention (less forgetting), while the Test (T) split measures predictive adaptability to future interactions. RAIE results are highlighted.}
  \label{tab:main-two-datasets}
  \begin{adjustbox}{max width=\textwidth, max totalheight=0.7\textheight}
  \small
  \setlength{\tabcolsep}{6pt}
  \renewcommand{\arraystretch}{0.9}
  \begin{tabular}{
      C{2.6cm}  % Backbone
      C{2.1cm}  % Variant
      C{1.2cm} C{1.2cm} C{1.2cm} C{1.2cm} % ML10M: O.R@10 O.N@10 T.R@10 T.N@10
      C{0.25cm} % spacer
      C{1.2cm} C{1.2cm} C{1.2cm} C{1.2cm}  % Yelp:   O.R@10 O.N@10 T.R@10 T.N@10
  }
    \toprule
    & & \multicolumn{4}{c}{\textbf{MovieLens-10M}} & & \multicolumn{4}{c}{\textbf{Yelp}} \\
    \cmidrule(lr){3-6} \cmidrule(lr){8-11}
    \textbf{Backbone} & \textbf{Variant}
    & \makecell{S.\\R@10} & \makecell{S.\\N@10} & \makecell{T.\\R@10} & \makecell{T.\\N@10}
    & & \makecell{S.\\R@10} & \makecell{S.\\N@10} & \makecell{T.\\R@10} & \makecell{T.\\N@10} \\
    \midrule
    \multirow{8}{*}{\textbf{BERT4Rec}}
      & Base             & 0.1311 & 0.0670 & 0.0440 & 0.0216 && 0.0210
 & 0.0104
 & 0.0094
 & 0.0047
 \\
      & +LoRA            & 0.1103 & 0.0558 & 0.0569 & 0.0286 && 0.0194
 & 0.0097
 & 0.0086
 & 0.0041
 \\
      & +LoRA+Replay     & 0.1248 & 0.0644 & 0.0563 & 0.0282 && 0.0202
 & 0.0101
 & 0.0088
 & 0.0043
 \\
      & +LoRA+LwF        & 0.1116 & 0.0563 & 0.0573 & 0.0288 && 0.0241
 & 0.0123
 & 0.0145
 & 0.0069
 \\
 & +LSAT        & 0.1579 & 0.0816 & 0.0561 & 0.0279 && 0.0325
 & \textbf{0.0160}
 & 0.0176
 & 0.0086
 \\
  & +MoLE        & 0.1618 & 0.0837 & 0.0467 & 0.0229 && \textbf{0.0327}
 & 0.0159
 & 0.0170
 & 0.0083
 \\
  & +E-BPR        & 0.1310 & 0.0669 & 0.0440 & 0.0215 && 0.0255
 & 0.0126
 & 0.0126
 & 0.0060
 \\
    \raierow
      & \textbf{+RAIE}   & \textbf{0.1795} & \textbf{0.0957} & \textbf{0.0870} & \textbf{0.0453} && 0.0257
 & 0.0126
 & \textbf{0.0195
} & \textbf{0.0095
} \\
    \midrule
    \multirow{8}{*}{\textbf{SASRec}}
      & Base             & 0.0703 & 0.0335 & 0.0234 & 0.0108 && 0.0230
 & 0.0112
 & 0.0099
 & 0.0049
 \\
      & +LoRA            & 0.0509 & 0.0236 & 0.0366 & 0.0172 && 0.0208

 & 0.0104
 & 0.0119
 & 0.0058
 \\
      & +LoRA+Replay     & 0.0624 & 0.0296 & 0.0322 & 0.0150 && 0.0214
 & 0.0104
 & 0.0119
 & 0.0059
 \\
      & +LoRA+LwF        & 0.0524 & 0.0246 & 0.0374 & 0.0176 && 0.0210
 & 0.0103
 & 0.0111
 & 0.0056
 \\
       & +LSAT        & 0.0625 & 0.0295 & 0.0348 & 0.0162 && 0.0217
 & 0.0106
 & 0.0122
 & 0.0057
 \\
        & +MoLE        & 0.0703 & 0.0335 & 0.0234 & 0.0107 && 0.0230
 & 0.0112
 & 0.0099
 & 0.0049
 \\
        & +E-BPR        & \textbf{0.0984} & \textbf{0.0460} & 0.0344 & 0.0166 && \textbf{0.0268}
 & \textbf{0.0127}
 & 0.0083
 & 0.0040
 \\
    \raierow
      & \textbf{+RAIE}   & 0.0686 & 0.0327 & \textbf{0.0449} & \textbf{0.0209} && 0.0251
 & 0.0123
 & \textbf{0.0125
} & \textbf{0.0062
} \\
    \midrule
    \multirow{8}{*}{\textbf{TiSASRec}}
      & Base             & 0.1093 & 0.0589 & 0.0342 & 0.0167 && 0.0693
 & 0.0321
 & 0.0123
 & 0.0055
 \\
      & +LoRA            & 0.1330 & 0.0686 & 0.0449 & 0.0220 && 0.0580
 & 0.0278
 & 0.0130
 & 0.0061
 \\
      & +LoRA+Replay     & 0.1226 & 0.0646 & 0.0467 & 0.0225 && 0.0572
 & 0.0274
 & 0.0130
 & 0.0062
 \\
      & +LoRA+LwF        & 0.1303 & 0.0675 & 0.0467 & 0.0225 && 0.0561
 & 0.0271
 & 0.0135
 & 0.0063
 \\
 & +LSAT        & 0.1118 & 0.0583 & 0.0359 & 0.0176 && 0.0721
 & 0.0333
 & 0.0136
 & 0.0059
 \\
  & +MoLE        & 0.1338 & 0.0695 & 0.0467 & 0.0227 && 0.0721
 & 0.0333
 & 0.0136
 & 0.0059
 \\
  & +E-BPR        & \textbf{0.1666} & \textbf{0.0870} & 0.0443 & 0.0218 && 0.0661
 & 0.0317
 & 0.0127
 & 0.0057
 \\
    \raierow
      & \textbf{+RAIE}   & 0.1646 & 0.0868 & \textbf{0.0483} & \textbf{0.0231} && \textbf{0.0833
} & \textbf{0.0381
} & \textbf{0.0135} & \textbf{0.0063
} \\
    \midrule
    \multirow{8}{*}{\textbf{openP5}}
      & Base             & 0.2390 & 0.1422 & 0.0356 & 0.0196 && 0.0123 & 0.0062 & 0.0051 & 0.0024 \\
      & +LoRA            & 0.1109     & 0.0570     & 0.0877     & 0.0470     && 0.0071 & 0.0033 & 0.0068 & 0.0033 \\
      & +LoRA+Replay     & 0.2675     & 0.1560     & 0.0887     & 0.0474     && 0.0105 & 0.0051 & 0.0079 & 0.0038 \\
      & +LoRA+LwF        & 0.2456     & 0.1488     & 0.0890     & 0.0477     && 0.0101 & 0.0048 & 0.0083 & 0.0040 \\
            & +LSAT        & 0.1107     & 0.0578     & 0.0572     & 0.0293     && 0.0003 & 0.0001 & 0.0002 & 0.0001 \\
                        & +MoLE        & 0.0384     & 0.0183     & 0.0579     & 0.0276     && 0.0043 & 0.0021 & 0.0043 & 0.0019 \\
                                    & +E-BPR        & 0.2556     & 0.1503     & 0.0867     & 0.0463     && 0.0111 & 0.0053 & 0.0076 & 0.0037 \\
    \raierow
      & \textbf{+RAIE}   & \textbf{0.2768} & \textbf{0.1686} & \textbf{0.0935} & \textbf{0.0486} && \textbf{0.0125} & \textbf{0.0062} & \textbf{0.0085} & \textbf{0.0043} \\
    \bottomrule
  \end{tabular}
  \end{adjustbox}
\end{table*}

\subsubsection{Baselines for Comparison.}
We benchmark RAIE against incremental learning plugins across various backbone models. All methods follow the same blocked incremental evaluation protocol.

\textbf{Sequence Recommendation Methods:}
{\textbf{BERT4Rec}}~\cite{sun2019bert4rec}: bidirectional Transformer with masked-item prediction. {\textbf{SASRec}}~\cite{kang2018self}: causal self‑attention for next‑item prediction. \textbf{TiSASRec}~\cite{5e2d653a3a55acc837436820}: self‑
attention enhanced with relative time intervals.

\textbf{LLM-Based Recommendation Methods:}
\textbf{OpenP5}~\cite{xu2024openp5}: unified LLM‑for‑Rec platform.

\textbf{Incremental plugins :}
\textbf{Replay}~\cite{599c7972601a182cd263ed19}: rehearsal via episodic memory. \textbf{LwF}~\cite{57a4e91aac44365e35c97dc2}: knowledge distillation to retain prior knowledge.

\textbf{LLM-based Parameter Efficient Tuning:}
\textbf{LSAT}~\cite{shi2024preliminary}: dual‑LoRA (long‑term stable + short‑term adaptive). \textbf{MoLE}~\cite{kong2024customizing}: mixture of LoRA experts with learned gating.

\textbf{Knowledge Editing based Recommendation Methods:}
\textbf{E-BPR}~\cite{lai2024better}: using novel editing bayesian personalized ranking loss for recommendation editing. 

\subsection{Overall Performance (RQ1)} Table~\ref{tab:main-two-datasets} reports Recall@10/NDCG@10 on the Set-up (S) split and the Test (T) split. We first observe that adding LoRA to the frozen backbone consistently improves T over the Base model, indicating that parameter-efficient fine-tuning enhances adaptability to preference drift; however, this global adaptation also degrades S, revealing partial forgetting of historical preferences. LoRA+Replay and LoRA+LwF mitigate this forgetting, recovering S while maintaining solid T performance, which suggests that rehearsal or consistency regularization helps retain prior knowledge, though gains on future data remain limited. LSAT and MoLE generally achieve stronger S (retention) than plain LoRA, but their improvements on T are modest and can be backbone-dependent or dataset-dependent, indicating limited generalization. E-BPR, which edits by retraining on mispredicted user–item behavior pairs, excels on S due to strong memorization of corrected interactions but generally lags on T because it lacks an explicit mechanism for modeling evolving intents and thus generalizes poorly under preference shift. Across all methods, scores are lower on Yelp than on MovieLens-10M owing to higher sparsity and noisier interactions, which weaken sequential signals and compress performance margins.

RAIE consistently achieves the best performance. It delivers the strongest T results across all backbones and datasets while maintaining competitive S performance, demonstrating a superior balance between retention and adaptation. We attribute these improvements to three key factors: (1) region-aware editing that confines updates to preference-coherent regions, reducing interference with preserved knowledge; (2) dynamic routing with localized LoRA adapters that activates parameters aligned with current intents for fine-grained adaptation; and (3) incremental, constrained updates that mitigate catastrophic forgetting and stabilize S.

\subsection{Ablation Study (RQ2)}
In order to validate the effectiveness of the proposed approach, we check the contribution of the main components of model to the final performance by comparing RAIE with the three variants:
\begin{itemize}[leftmargin=*]
\item w/o KR: removing the Knowledge Region construction. comparing a single global adapter with multiple regional adapters.

\item w/o KE: disabling the Region-Aware Editing. Evaluate its effectiveness for handling preference drift.

\item w/o All: removing the Knowledge Region Construction and the Region-Aware Editing.
\end{itemize}

\begin{figure}[t]
  \centering
  \begin{subfigure}{0.49\columnwidth}
    \centering
    \includegraphics[width=\linewidth]{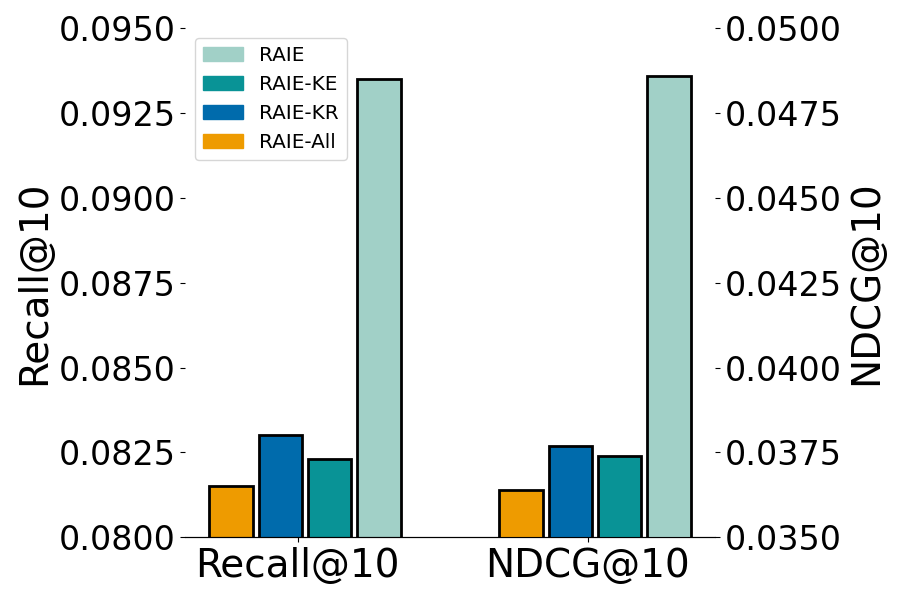}
    \caption{MovieLens 10M}
    \label{hyper:movieabl}
  \end{subfigure}
  \hfill
  \begin{subfigure}{0.49\columnwidth}
    \centering
    \includegraphics[width=\linewidth]{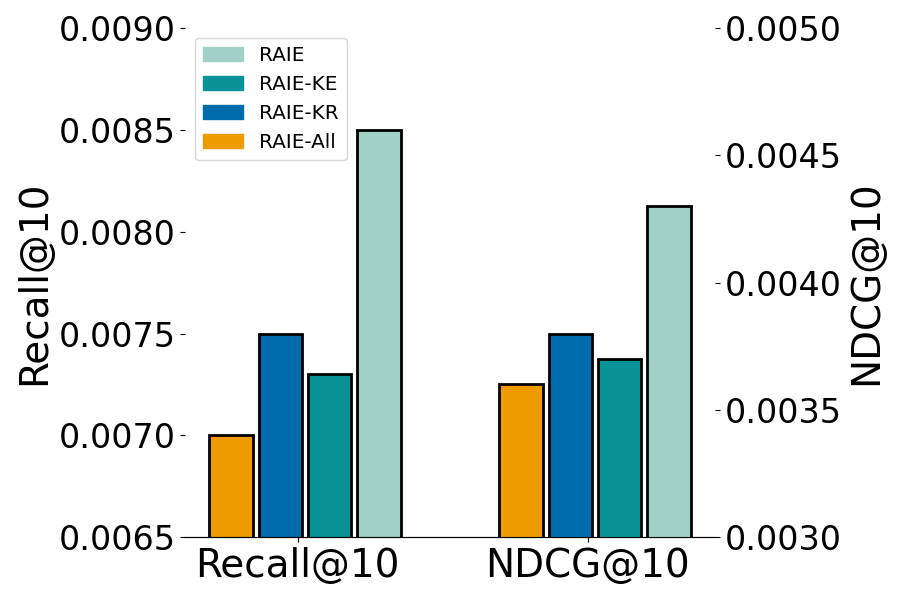}
    \caption{Yelp}
    \label{hyper:yelpabl}
  \end{subfigure}

  \caption{The ablation study of RAIE.}
  \label{fig:ablation}
\end{figure}

The experimental results are reported in Figure~\ref{fig:ablation}, we could summarize the following observations: (i) The proposed Region-Aware Editing contributes the most to performance enhancement. It is crucial to effectively update user preferences; moreover, region-aware editing effectively addresses preference drift. By editing the original preference representation to adapt to evolving interests, recommendation performance can be substantially improved. (ii) The variant ‘‘w/o KR’’ confronts conspicuous performance decay in recommendation. This indicates that constructing user preference knowledge regions effectively separates a user’s different preferences. Compared with a single global adapter, multiple local adapters can capture the user’s current preference from the perspective of each region and update preferences locally within that region. (iii) After removing both Knowledge Region Construction and Region-Aware Editing, the model’s performance drops substantially, indicating that these two modules are both important and effective. Combining them further improves recommendation performance.

\subsection{Hyperparameter Sensitivity (RQ3)} 
\begin{figure}[t]
  \centering
  \begin{subfigure}{0.49\columnwidth}
    \centering
    \includegraphics[width=\linewidth]{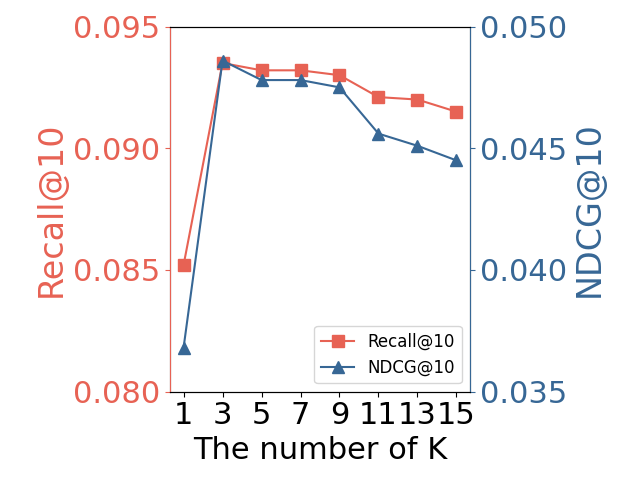}
    \caption{MovieLens 10M}
    \label{hyper:movie}
  \end{subfigure}
  \hfill
  \begin{subfigure}{0.49\columnwidth}
    \centering
    \includegraphics[width=\linewidth]{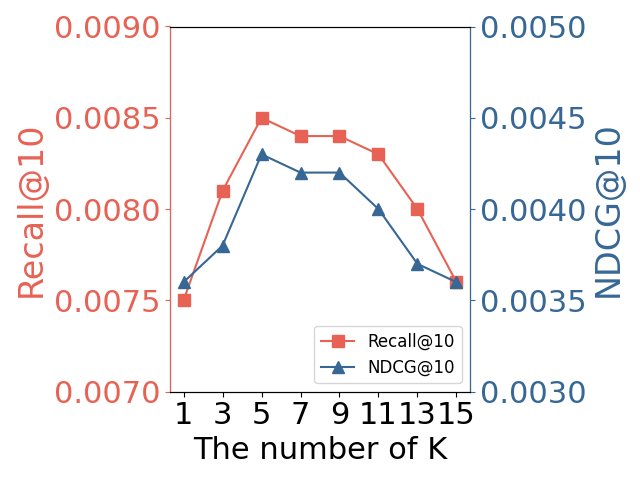}
    \caption{Yelp}
    \label{hyper:yelp}
  \end{subfigure}

  \caption{Impact of region number K.}
  \label{fig:k number}
\end{figure}

To analyze the effect of the number of regions K, we vary the K range and show the performance variation curves on both datasets in Figure~\ref{fig:k number}. We found that: when there are few regions, the overall performance of model is poor. This again emphasizes the benefit of exploring multiple user preference regions. However, as the number of regions increases, the model performance shows a decreasing trend. This is because an increase in the number of regions gradually reduces the amount of useful information that each knowledge regions carries, and instead introduces more noise information, which impairs the model’s performance. RAIE performs best on MovieLens-10M and Yelp with K = 3 and 5, respectively, which we attribute to the differences between the two datasets.

\subsection{Interpretability \& Visualization (RQ4)} 
\begin{figure}[t]
  \centering

  % Row 1: MovieLens 10M
  \textbf{MovieLens 10M}\par\smallskip
  \begin{subfigure}{0.32\columnwidth}
    \centering
    \includegraphics[width=\linewidth]{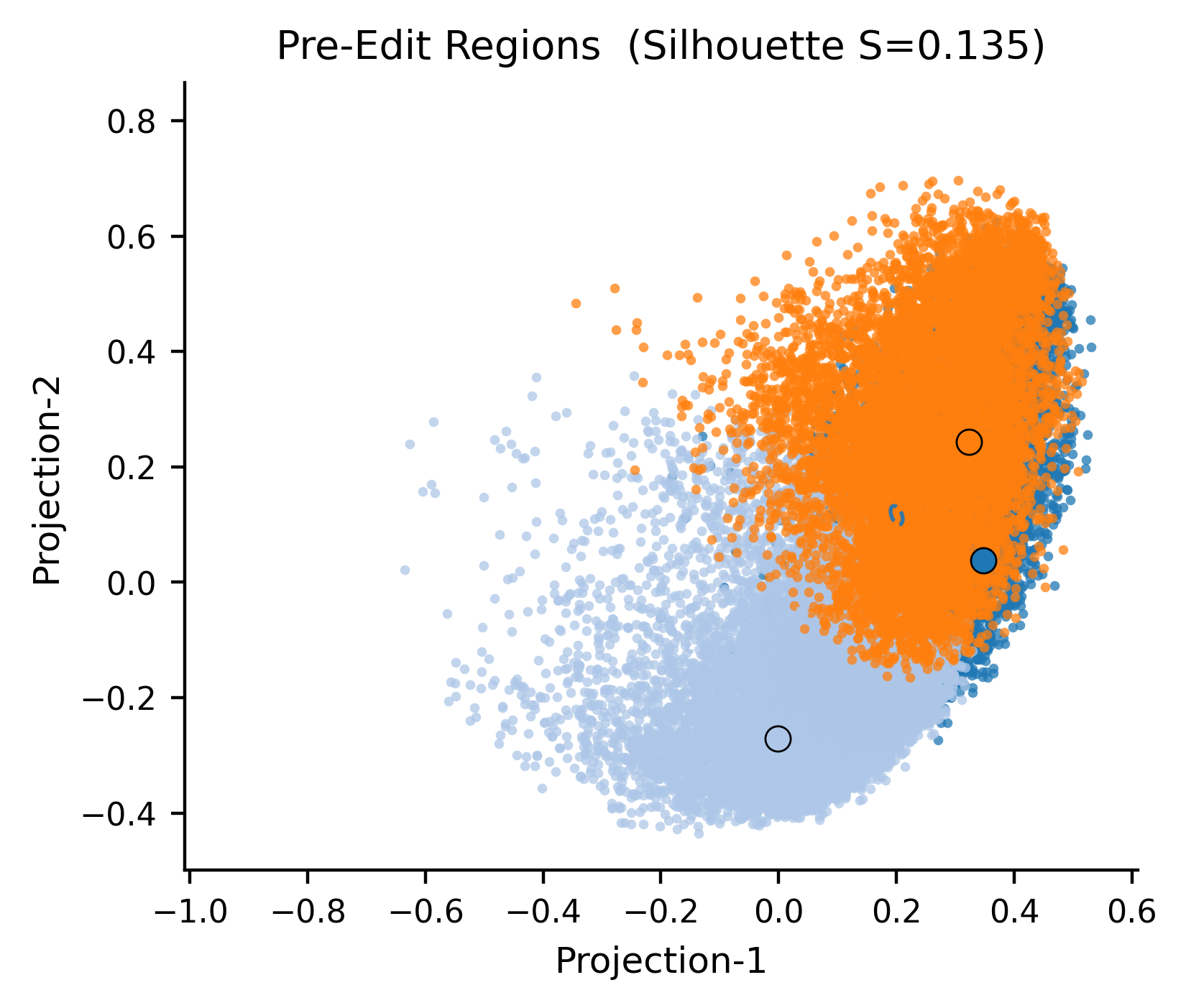}
    \caption{Pre-Edit Regions}
    \label{fig:movie:pre}
  \end{subfigure}
  \hfill
  \begin{subfigure}{0.32\columnwidth}
    \centering
    \includegraphics[width=\linewidth]{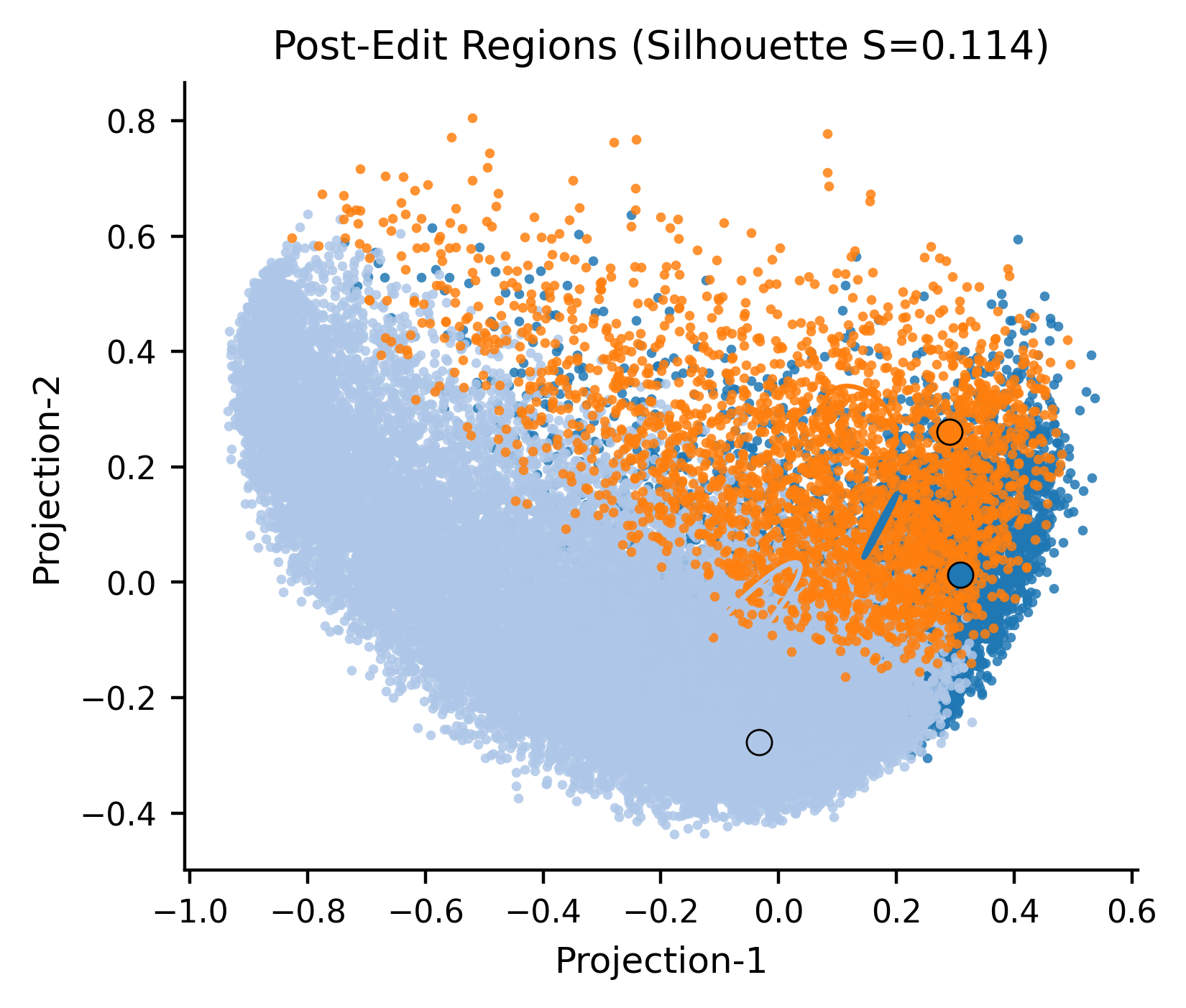}
    \caption{Post-Edit Regions}
    \label{fig:movie:post}
  \end{subfigure}
  \hfill
  \begin{subfigure}{0.32\columnwidth}
    \centering
    \includegraphics[width=\linewidth]{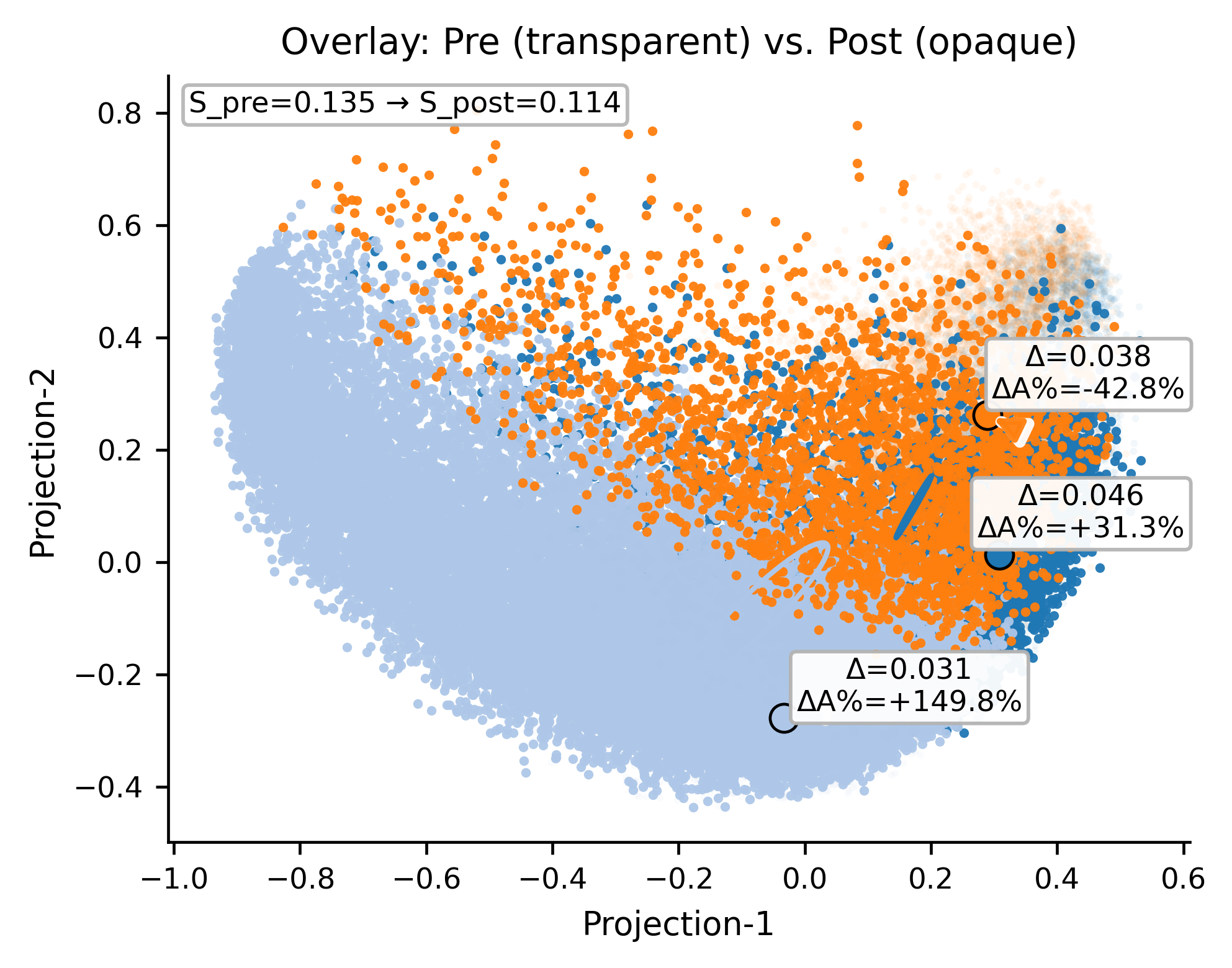}
    \caption{Overlay}
    \label{fig:movie:overlay}
  \end{subfigure}

  \medskip

  % Row 2: Yelp
  \textbf{Yelp}\par\smallskip
  \begin{subfigure}{0.32\columnwidth}
    \centering
    \includegraphics[width=\linewidth]{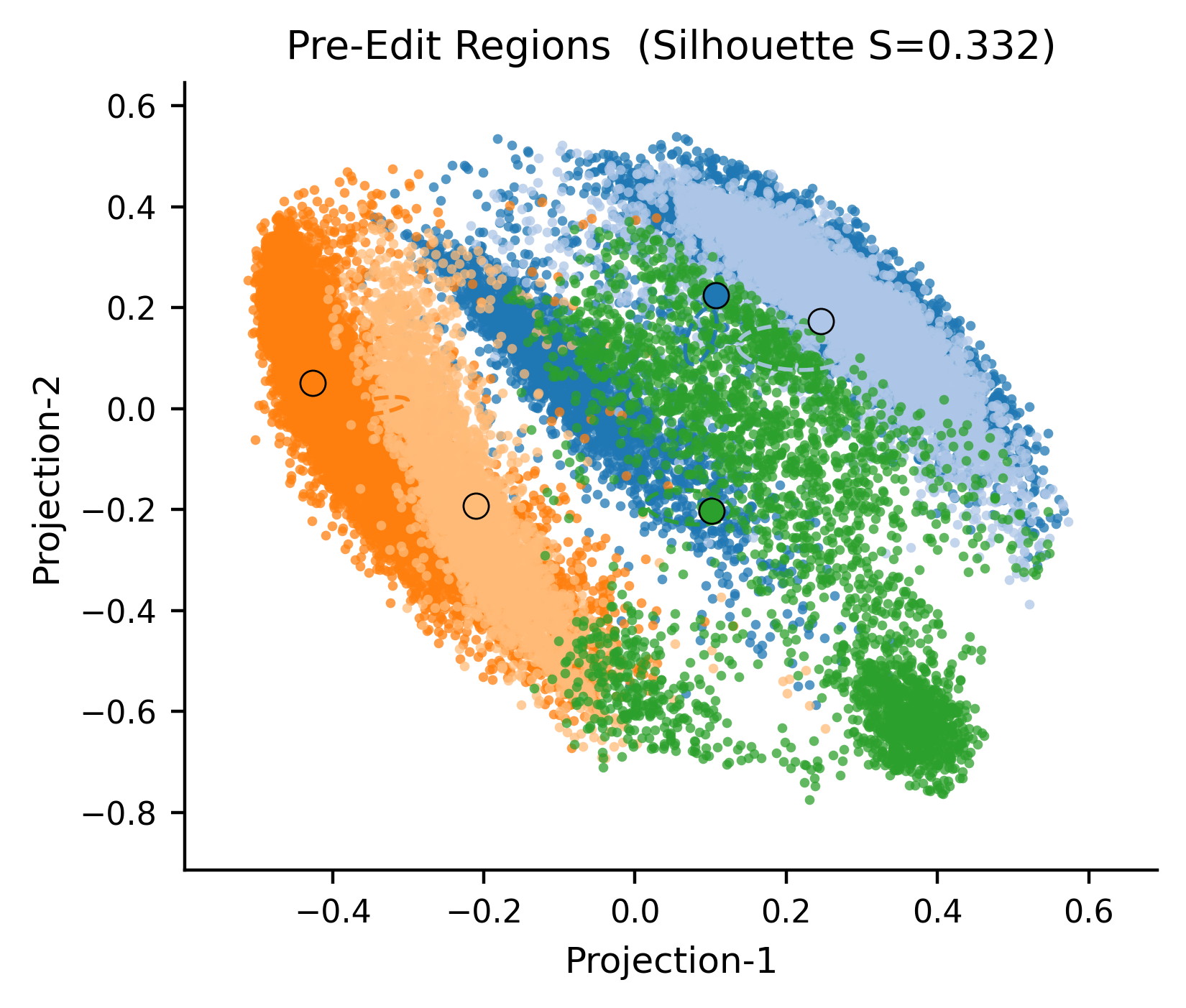}
    \caption{Pre-Edit Regions}
    \label{fig:yelp:pre}
  \end{subfigure}
  \hfill
  \begin{subfigure}{0.32\columnwidth}
    \centering
    \includegraphics[width=\linewidth]{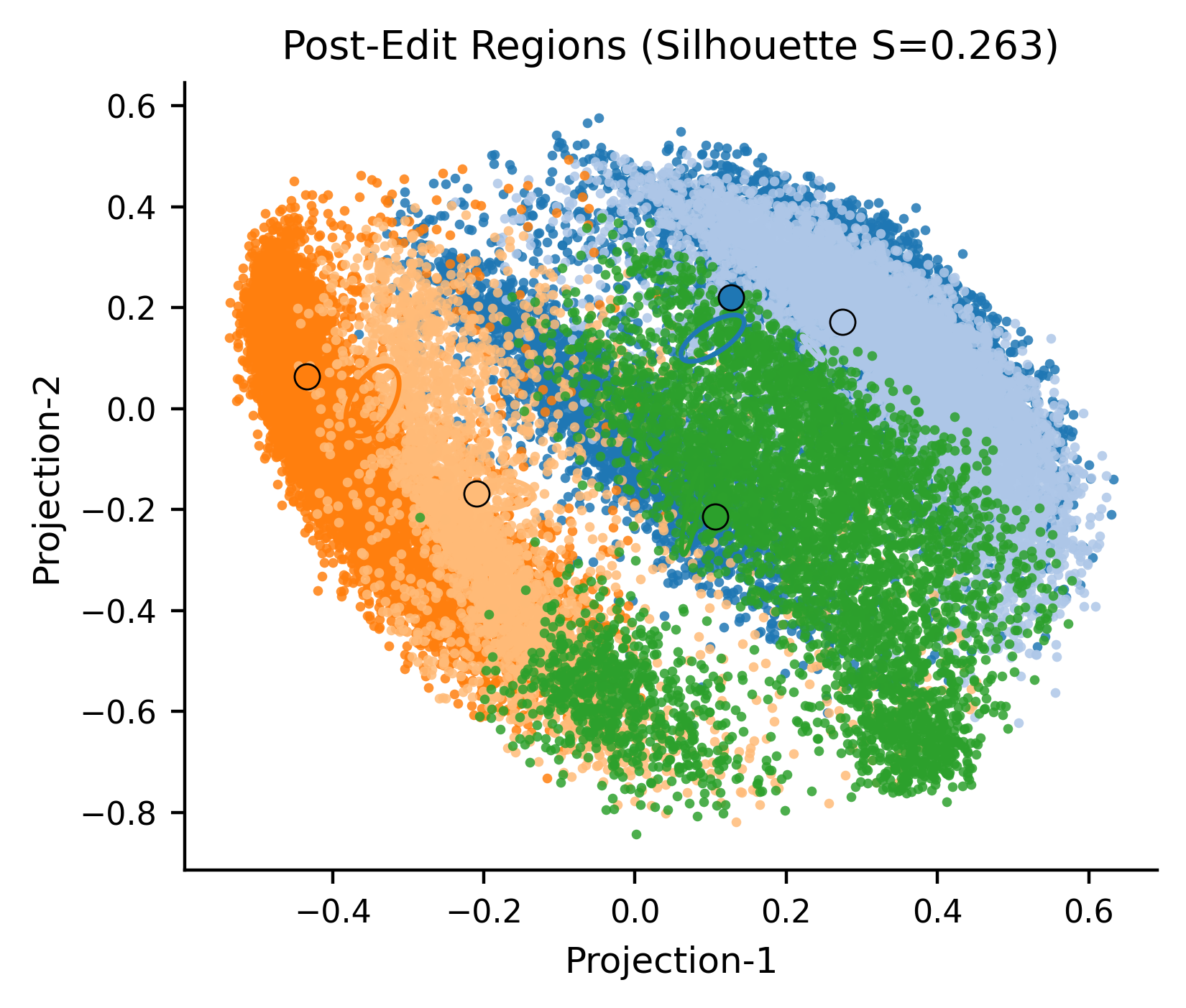}
    \caption{Post-Edit Regions}
    \label{fig:yelp:post}
  \end{subfigure}
  \hfill
  \begin{subfigure}{0.32\columnwidth}
    \centering
    \includegraphics[width=\linewidth]{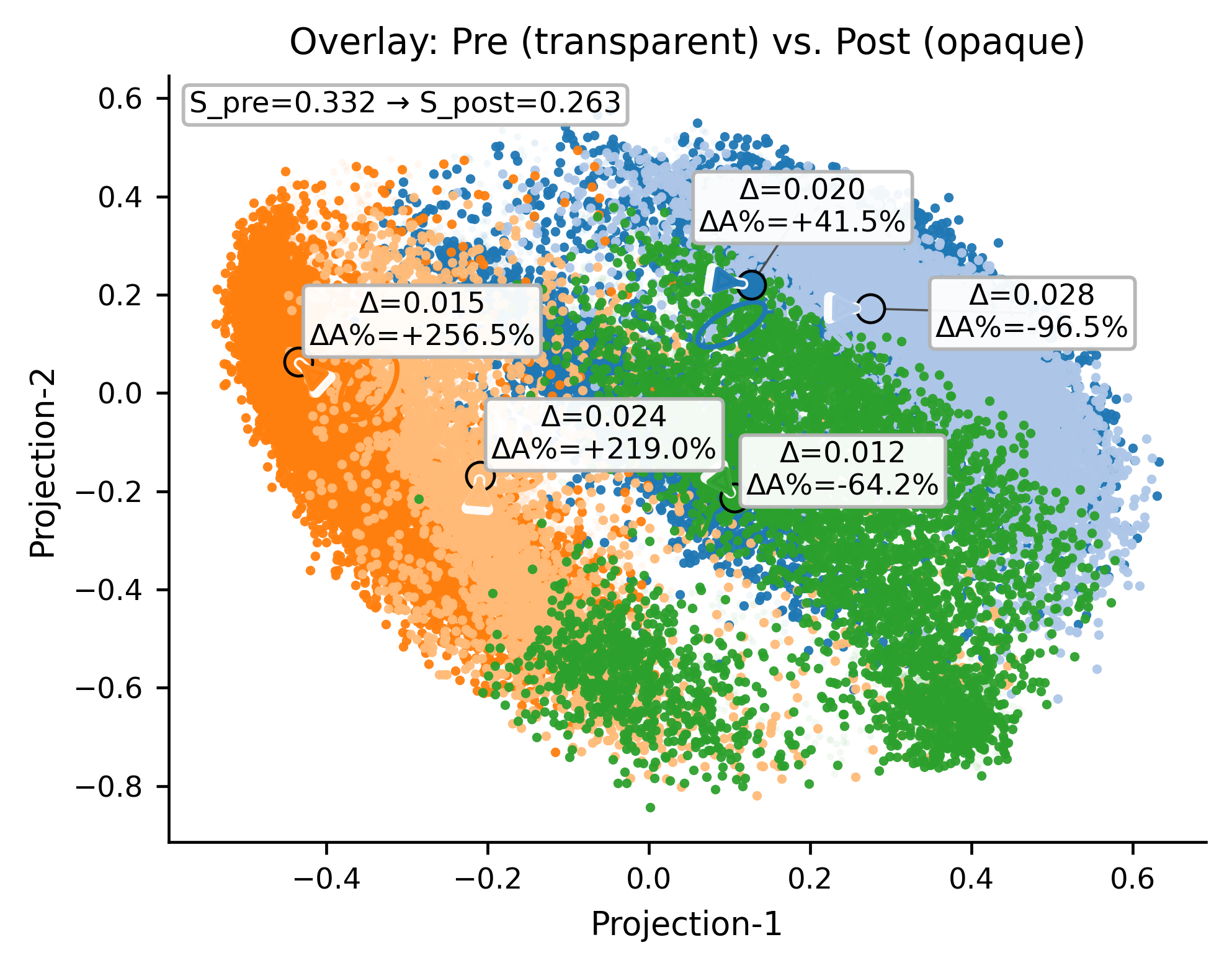}
    \caption{Overlay}
    \label{fig:yelp:overlay}
  \end{subfigure}

  \caption{The visualization of region distributions.}
  \label{fig:vis_all}
\end{figure}

Finally, we examine interpretability by visualizing the knowledge regions before and after editing. Figure~\ref{fig:vis_all} visualizes pre-edit and post-edit knowledge regions for MovieLens and Yelp (with overlays) and annotates centroid displacement ($\Delta$), relative boundary–area change ($\Delta$A\%), and region separability (S). Across both datasets, the overlays show that editing preserves the global partition while inducing local, targeted adjustments: centroids shift modestly and boundaries expand primarily near interfaces, reallocating coverage without re-partitioning the space. The separability metric changes in line with these boundary refinements, indicating comparable inter-region distinction after editing. Comparing domains, MovieLens exhibits larger $\Delta$ with moderate ($\Delta$A\%), whereas Yelp shows smaller centroid movement but substantially larger |$\Delta$A\%|, indicating stronger reallocation of boundary coverage. The separability metric $S$ tends to decrease slightly after editing on both datasets, reflecting smoother inter-region transitions rather than new partitioning. Overall, RAIE adapts regional geometry to evolving preferences while maintaining an interpretable structure, consistent with the observed gains in recommendation quality.

\section{Conclusion}
We introduce RAIE, a plug‑in, parameter‑efficient framework for region‑aware incremental preference editing. Its modular design is backbone‑agnostic and robust under diverse drift patterns. While effective, RAIE currently relies on spherical $k$‑means with a fixed $K$, making performance sensitive to this choice. Future work will focus on adaptive region discovery that learns both the number of regions and their boundaries, enhancing robustness.

\section{Acknowledgements}
This work is supported and sponsored by the Guangdong Basic and Applied Basic Research Foundation (2023A1515012848), National Natural Science Foundation of China (Nos. 62572410, 62576371 and T2122020), and CCF-DiDi GAIA Collaborative Research Funds.

\bibliographystyle{ACM-Reference-Format}
\bibliography{References}

@String{Chelsea = "Chelsea" }

@article{zhang2024comprehensive,
  title={A comprehensive study of knowledge editing for large language models},
  author={Zhang, Ningyu and Yao, Yunzhi and Tian, Bozhong and Wang, Peng and Deng, Shumin and Wang, Mengru and Xi, Zekun and Mao, Shengyu and Zhang, Jintian and Ni, Yuansheng and others},
  journal={arXiv preprint arXiv:2401.01286},
  year={2024}
}

@article{wei2025stability,
  title={Stability-aware Preference Modeling for Sequential Recommendation},
  author={Wei, Chaoyong and Wenjun, Jiang and Li, Kenli and Wu, Jie},
  journal={ACM Transactions on the Web},
  year={2025},
  publisher={ACM New York, NY}
}

@article{huang2025towards,
  title={Towards agentic recommender systems in the era of multimodal large language models},
  author={Huang, Chengkai and Wu, Junda and Xia, Yu and Yu, Zixu and Wang, Ruhan and Yu, Tong and Zhang, Ruiyi and Rossi, Ryan A and Kveton, Branislav and Zhou, Dongruo and others},
  journal={arXiv preprint arXiv:2503.16734},
  year={2025}
}

@inproceedings{lin2023self,
  title={A self-correcting sequential recommender},
  author={Lin, Yujie and Wang, Chenyang and Chen, Zhumin and Ren, Zhaochun and Xin, Xin and Yan, Qiang and de Rijke, Maarten and Cheng, Xiuzhen and Ren, Pengjie},
  booktitle={Proceedings of the ACM Web Conference 2023},
  pages={1283--1293},
  year={2023}
}

@article{hu2022lora,
  title={Lora: Low-rank adaptation of large language models.},
  author={Hu, Edward J and Shen, Yelong and Wallis, Phillip and Allen-Zhu, Zeyuan and Li, Yuanzhi and Wang, Shean and Wang, Lu and Chen, Weizhu and others},
  journal={ICLR},
  volume={1},
  number={2},
  pages={3},
  year={2022}
}

@inproceedings{Kang2018SASRec,
  title={Self-Attentive Sequential Recommendation},
  author={Kang, Wang-Cheng and McAuley, Julian},
  booktitle={2018 IEEE International Conference on Data Mining (ICDM)},
  pages={197--206},
  year={2018}
}

@article{dhillon:modha:mlj01,
      AUTHOR = {Dhillon, I. S. and Modha, D. S.},
      TITLE = { Concept decompositions for large sparse text data using clustering},
      JOURNAL = {Machine Learning},
      YEAR = {2001},
      MONTH = {Jan},
      VOLUME = {42},
      NUMBER = {1},
      PAGES = {143--175} }

@inproceedings{houlsby2019parameter,
  title={Parameter-efficient transfer learning for NLP},
  author={Houlsby, Neil and Giurgiu, Andrei and Jastrzebski, Stanislaw and Morrone, Bruna and De Laroussilhe, Quentin and Gesmundo, Andrea and Attariyan, Mona and Gelly, Sylvain},
  booktitle={International conference on machine learning},
  pages={2790--2799},
  year={2019},
  organization={PMLR}
}

@article{li2021prefix,
  title={Prefix-tuning: Optimizing continuous prompts for generation},
  author={Li, Xiang Lisa and Liang, Percy},
  journal={arXiv preprint arXiv:2101.00190},
  year={2021}
}

@article{meng2022locating,
  title={Locating and editing factual associations in gpt},
  author={Meng, Kevin and Bau, David and Andonian, Alex and Belinkov, Yonatan},
  journal={Advances in neural information processing systems},
  volume={35},
  pages={17359--17372},
  year={2022}
}

@article{meng2022mass,
  title={Mass-editing memory in a transformer},
  author={Meng, Kevin and Sharma, Arnab Sen and Andonian, Alex and Belinkov, Yonatan and Bau, David},
  journal={arXiv preprint arXiv:2210.07229},
  year={2022}
}

@inproceedings{mitchell2022memory,
  title={Memory-based model editing at scale},
  author={Mitchell, Eric and Lin, Charles and Bosselut, Antoine and Manning, Christopher D and Finn, Chelsea},
  booktitle={International Conference on Machine Learning},
  pages={15817--15831},
  year={2022},
  organization={PMLR}
}

@inproceedings{kang2018self,
  title={Self-attentive sequential recommendation},
  author={Kang, Wang-Cheng and McAuley, Julian},
  booktitle={2018 IEEE international conference on data mining (ICDM)},
  pages={197--206},
  year={2018},
  organization={IEEE}
}

@article{mitchell2021fast,
  title={Fast model editing at scale},
  author={Mitchell, Eric and Lin, Charles and Bosselut, Antoine and Finn, Chelsea and Manning, Christopher D},
  journal={arXiv preprint arXiv:2110.11309},
  year={2021}
}

@inproceedings{shi2024preliminary,
  title={Preliminary study on incremental learning for large language model-based recommender systems},
  author={Shi, Tianhao and Zhang, Yang and Xu, Zhijian and Chen, Chong and Feng, Fuli and He, Xiangnan and Tian, Qi},
  booktitle={Proceedings of the 33rd ACM International Conference on Information and Knowledge Management},
  pages={4051--4055},
  year={2024}
}

@article{kong2024customizing,
  title={Customizing language models with instance-wise lora for sequential recommendation},
  author={Kong, Xiaoyu and Wu, Jiancan and Zhang, An and Sheng, Leheng and Lin, Hui and Wang, Xiang and He, Xiangnan},
  journal={Advances in Neural Information Processing Systems},
  volume={37},
  pages={113072--113095},
  year={2024}
}

@inproceedings{koren2009collaborative,
  title={Collaborative filtering with temporal dynamics},
  author={Koren, Yehuda},
  booktitle={Proceedings of the 15th ACM SIGKDD international conference on Knowledge discovery and data mining},
  pages={447--456},
  year={2009}
}

@inproceedings{sun2019bert4rec,
  title={BERT4Rec: Sequential recommendation with bidirectional encoder representations from transformer},
  author={Sun, Fei and Liu, Jun and Wu, Jian and Pei, Changhua and Lin, Xiao and Ou, Wenwu and Jiang, Peng},
  booktitle={Proceedings of the 28th ACM international conference on information and knowledge management},
  pages={1441--1450},
  year={2019}
}

@article{shin2017continual,
  title={Continual learning with deep generative replay},
  author={Shin, Hanul and Lee, Jung Kwon and Kim, Jaehong and Kim, Jiwon},
  journal={Advances in neural information processing systems},
  volume={30},
  year={2017}
}

@inproceedings{xuhong2018explicit,
  title     = {Explicit Inductive Bias for Transfer Learning with Convolutional Networks},
  author    = {Li, Xuhong and Grandvalet, Yves and Davoine, Franck},
  booktitle = {Proceedings of the 35th International Conference on Machine Learning},
  series    = {Proceedings of Machine Learning Research},
  volume    = {80},
  pages     = {2825--2834},
  year      = {2018},
  publisher = {PMLR},
}

@article{parisi2019continual,
  title={Continual lifelong learning with neural networks: A review},
  author={Parisi, German I and Kemker, Ronald and Part, Jose L and Kanan, Christopher and Wermter, Stefan},
  journal={Neural networks},
  volume={113},
  pages={54--71},
  year={2019},
  publisher={Elsevier}
}

@article{kirkpatrick2017overcoming,
  title={Overcoming catastrophic forgetting in neural networks},
  author={Kirkpatrick, James and Pascanu, Razvan and Rabinowitz, Neil and Veness, Joel and Desjardins, Guillaume and Rusu, Andrei A and Milan, Kieran and Quan, John and Ramalho, Tiago and Grabska-Barwinska, Agnieszka and others},
  journal={Proceedings of the national academy of sciences},
  volume={114},
  number={13},
  pages={3521--3526},
  year={2017},
  publisher={National Academy of Sciences}
}

@article{rajput2023recommender,
  title={Recommender systems with generative retrieval},
  author={Rajput, Shashank and Mehta, Nikhil and Singh, Anima and Hulikal Keshavan, Raghunandan and Vu, Trung and Heldt, Lukasz and Hong, Lichan and Tay, Yi and Tran, Vinh and Samost, Jonah and others},
  journal={Advances in Neural Information Processing Systems},
  volume={36},
  pages={10299--10315},
  year={2023}
}

@inproceedings{yang2022knowledge,
  title={Knowledge graph contrastive learning for recommendation},
  author={Yang, Yuhao and Huang, Chao and Xia, Lianghao and Li, Chenliang},
  booktitle={Proceedings of the 45th international ACM SIGIR conference on research and development in information retrieval},
  pages={1434--1443},
  year={2022}
}

@article{zeng2025knowledge,
  title={Knowledge-driven hierarchical intents modeling for recommendation},
  author={Zeng, Jin and Wang, Nan and Li, Jinbao},
  journal={Expert Systems with Applications},
  volume={259},
  pages={125361},
  year={2025},
  publisher={Elsevier}
}

@article{liu2023llmrec,
  title={Llmrec: Benchmarking large language models on recommendation task},
  author={Liu, Junling and Liu, Chao and Zhou, Peilin and Ye, Qichen and Chong, Dading and Zhou, Kang and Xie, Yueqi and Cao, Yuwei and Wang, Shoujin and You, Chenyu and others},
  journal={arXiv preprint arXiv:2308.12241},
  year={2023}
}

@article{lewis2020retrieval,
  title={Retrieval-augmented generation for knowledge-intensive nlp tasks},
  author={Lewis, Patrick and Perez, Ethan and Piktus, Aleksandra and Petroni, Fabio and Karpukhin, Vladimir and Goyal, Naman and K{\"u}ttler, Heinrich and Lewis, Mike and Yih, Wen-tau and Rockt{\"a}schel, Tim and others},
  journal={Advances in neural information processing systems},
  volume={33},
  pages={9459--9474},
  year={2020}
}

@inproceedings{bao2023tallrec,
  title={Tallrec: An effective and efficient tuning framework to align large language model with recommendation},
  author={Bao, Keqin and Zhang, Jizhi and Zhang, Yang and Wang, Wenjie and Feng, Fuli and He, Xiangnan},
  booktitle={Proceedings of the 17th ACM conference on recommender systems},
  pages={1007--1014},
  year={2023}
}

@article{luo2025recranker,
  title={Recranker: Instruction tuning large language model as ranker for top-k recommendation},
  author={Luo, Sichun and He, Bowei and Zhao, Haohan and Shao, Wei and Qi, Yanlin and Huang, Yinya and Zhou, Aojun and Yao, Yuxuan and Li, Zongpeng and Xiao, Yuanzhang and others},
  journal={ACM Transactions on Information Systems},
  volume={43},
  number={5},
  pages={1--31},
  year={2025},
  publisher={ACM New York, NY}
}

@article{loshchilov2017decoupled,
  title={Decoupled weight decay regularization},
  author={Loshchilov, Ilya and Hutter, Frank},
  journal={arXiv preprint arXiv:1711.05101},
  year={2017}
}

@article{xu2024openp5,	author={Shuyuan Xu and Wenyue Hua and Yongfeng Zhang},	title={OpenP5: an Open-Source Platform for Developing, Training, and Evaluating LLM-based Recommender Systems},	year=2024,}

@article{klasson2022learn,
  title={Learn the time to learn: Replay scheduling in continual learning},
  author={Klasson, Marcus and Kjellstr{\"o}m, Hedvig and Zhang, Cheng},
  journal={arXiv preprint arXiv:2209.08660},
  year={2022}
}

@article{yoo2025continual,
  title={Continual Recommender Systems},
  author={Yoo, Hyunsik and Kang, SeongKu and Tong, Hanghang},
  journal={arXiv preprint arXiv:2507.03861},
  year={2025}
}

@inproceedings{li2025elder,
  title={ELDER: Enhancing Lifelong Model Editing with Mixture-of-LoRA},
  author={Li, Jiaang and Wang, Quan and Wang, Zhongnan and Zhang, Yongdong and Mao, Zhendong},
  booktitle={Proceedings of the AAAI Conference on Artificial Intelligence},
  volume={39},
  number={23},
  pages={24440--24448},
  year={2025}
}

@article{lai2024better,
  title={Better Late Than Never: Formulating and Benchmarking Recommendation Editing},
  author={Lai, Chengyu and Zhou, Sheng and Jiang, Zhimeng and Tan, Qiaoyu and Bei, Yuanchen and Chen, Jiawei and Zhang, Ningyu and Bu, Jiajun},
  journal={arXiv preprint arXiv:2406.04553},
  year={2024}
}

@inproceedings{geng2022recommendation,
  title={Recommendation as language processing (rlp): A unified pretrain, personalized prompt \& predict paradigm (p5)},
  author={Geng, Shijie and Liu, Shuchang and Fu, Zuohui and Ge, Yingqiang and Zhang, Yongfeng},
  booktitle={Proceedings of the 16th ACM conference on recommender systems},
  pages={299--315},
  year={2022}
}

@article{rozner2024knowledge,
  title={Knowledge editing in language models via adapted direct preference optimization},
  author={Rozner, Amit and Battash, Barak and Wolf, Lior and Lindenbaum, Ofir},
  journal={arXiv preprint arXiv:2406.09920},
  year={2024}
}

@article{ding2023parameter,
  title={Parameter-efficient fine-tuning of large-scale pre-trained language models},
  author={Ding, Ning and Qin, Yujia and Yang, Guang and Wei, Fuchao and Yang, Zonghan and Su, Yusheng and Hu, Shengding and Chen, Yulin and Chan, Chi-Min and Chen, Weize and others},
  journal={Nature machine intelligence},
  volume={5},
  number={3},
  pages={220--235},
  year={2023},
  publisher={Nature Publishing Group UK London}
}

@article{wu2024mixture,
  title={Mixture of lora experts},
  author={Wu, Xun and Huang, Shaohan and Wei, Furu},
  journal={arXiv preprint arXiv:2404.13628},
  year={2024}
}

@article{5e2d653a3a55acc837436820,	author={Jiacheng Li and Yujie Wang and Julian McAuley},	pages={322-330},	title={Time Interval Aware Self-Attention for Sequential Recommendation},	year=2020,}

@article{599c7972601a182cd263ed19,	author={Hanul Shin and Jung Kwon Lee and Jaehong Kim and Jiwon Kim},	title={Continual Learning with Deep Generative Replay.},	volume=30,	year=2017,}

@article{57a4e91aac44365e35c97dc2,	author={Zhizhong Li and Derek Hoiem},	pages={614-629},	title={Learning Without Forgetting.},	volume=9908,	year=2016,}
\end{document}